\documentclass{iopart}
\usepackage{latexsym}
\usepackage{graphicx}
\usepackage{ulem}
\usepackage{subfigure}
\def\Deg{\hbox{$^\circ$}}

\begin{document}

\title[CRTNT sensitivity]{Simulation of Cosmic Ray Tau Neutrino Telescope
(CRTNT) Experiment}

\author{J. L. Liu$^1$, S. S. Zhang$^1$,  Z. Cao$^1$, H. H. He$^1$,
M. A. Huang$^2$, T. C. Liu$^3$, G. Xiao$^1$, M. Zha$^1$, B. K.
Zhang$^1$, Y. X. Bai$^1$, Y. Zhang$^1$}
\address{$^1$ Key Laboratory of Astroparticle Physics, Institute of High Energy Physics, Beijing 100049, China}
\address{$^2$ Department of Energy and Resources, National United University,
  1 Lienda, Miao-Li, 36003, Taiwan}
\address{$^3$ Institute of Physics, National Chiao-Tung University,
  1001 Ta-Hsueh Rd, Hsinchu 300 Taiwan}

\ead{liujl@ihep.ac.cn}

\begin{abstract}
A $\tau$ lepton can be produced in a charged current interaction
by cosmic ray tau neutrino with material inside a mountain. If it
escapes from the mountain, it will decay and initiate a shower in
the air, which can be detected by an air shower
fluorescence/Cherenkov light detector. Designed according to such
a principle, the Cosmic Ray Tau Neutrino Telescope (CRTNT)
experiment, located at the foothill of Mt. Balikun in Xinjiang,
China, will search for very high-energy cosmic $\tau$ neutrinos
from energetic astrophysical sources by detecting those showers.
This paper describes a Monte Carlo simulation for a detection of
$\tau $ events by the CRTNT experiment. Ultra-high-energy cosmic
ray events are also simulated to estimate the potential
contamination. With the CRTNT experiment composed of four detector
stations, each covering $64\Deg \times 14\Deg$ field of view, the
expected event rates are 28.6, 21.9 and 4.7 per year assuming AGN
neutrino flux according to Semikoz et. al. 2004, MPR AGN jet model
and SDSS AGN core model, respectively. Null detection of such
$\tau $ event by the CRTNT experiment in one year could set 90{\%}
C.L. upper limit at $19.9 (eV\cdot cm^{-2}\cdot s^{-1}\cdot
sr^{-1})$ for $E^{-2}$ neutrino spectrum.
\end{abstract}


\submitto{\JPG}
\maketitle

\section{Introduction}

The origin of cosmic rays (CR) above $10^{15}$ eV (1 PeV) has been
a long-standing problem. Observation of neutral particles, which
can be directly traced back to the sources, is a unique way to
search for point sources of CRs under $10^{19}$ eV. There are many
energetic astrophysical sources such as Gamma Ray Bursts (GRB) or
Active Galactic Nuclei (AGN). Those objects could accelerate
particles to ultra-high energy and produce CRs, gamma rays, and
neutrinos \cite{Learned00, Waxman03}. For photons at $10^{14}$ eV
to $10^{16}$ eV, their attenuation length for interaction with
cosmological microwave background photons are approximately 10
kpc, roughly the distance from the solar system to the galactic
center. Neutrinos are the only choice for exploring remote CR
sources. Observation of ultra-high-energy neutrinos is not only an
important proof of their existence but also an orthogonal tool,
besides conventional astronomical tools such as optical and radio
observation, for studying GRBs or AGNs. However, because of very
weak interaction, neutrino detection requires a huge volume of
detector medium. Conventional way of neutrino detection is to
build or locate a huge volume of target and bury detectors inside
so that charged particles as products of interaction between
incident neutrinos and target material can be detected. As an
example, the IceCube experiment, located near the geographic South
Pole in Antarctica, uses natural ice approximately 2km under the
surface as the target and strings of UV sensitive optical modules
buried in the ice to detect Cherenkov light generated by muons and
electrons that are produced in interaction of incident
neutrinos\cite{icecube07}. Working in the similar principle, the
Antares experiment uses natural water in Mediterranean sea instead
of ice and sinks strings of similar optical modules down to the
sea bed to detect Cherenkov photons\cite{antares05}. The main
advantage of underwater/ice neutrino detectors is in the ability
to collect muons from large distances since muon range is much
longer compared to electron or tau range. For electrons generated
by e-neutrino, cascade in the sensitive area of the detector can
be reconstructed. For $\tau$ neutrinos in certain energy range,
both interaction vertex where $\tau$ lepton is produced and $\tau$
decay vertex might be contained in the sensitive area so that
yields a clear signature of "double bang". Therefore, the detector
is sensitive to all species of incident neutrinos with somewhat
clear signature for different species. Due to a limited
attenuation length of Cherenkov light, spacing between optical
modules must be sufficiently small in order to collect sufficient
Cherenkov photons that enables reasonable reconstruction of tracks
or showers inside the detector. This is a disadvantage of this
type of detector in terms of costs of construction. It is
substantial in the case of neutrino observation with very low
statistics, e.g. typically tens of events per year for a scale of
experiments such as IceCube. To seek some more economic ways to
build larger detectors for statistically significant measurements
is in fact essential. Sometimes, one has to sacrifice some
performance of the detector to fulfill the goal. One of many ways
to explore economic detection of ultra-high-energy neutrinos is to
separate the detector volume from the target volume in which the
incident neutrino interacts. By doing so, the target volume can be
as large as a piece of mountain or even a part of the earth shell
without putting any detector inside, and the detector volume can
be as large as the nearby atmosphere, such as that for the HiRes
experiment\cite{HiRes_nue} and the Pierre Auger Observatory (PAO)
experiment\cite{auger05}, both are primarily designed for
detecting cosmic rays above 1EeV. An obvious disadvantage is that
the interaction vertex is no longer in the scope of the detector.
A potential difficulty might be that the observation could be
contaminated by the cosmic ray background. There exists a solution
to avoid the cosmic ray background by slightly reduce the detector
volume with a more dedicated design of configuration that uses the
fluorescence/Cherenkov telescopes, such as HiRes/Auger telescopes,
to watch a volume of a few kilometers behind a mountain and keep
the field of view of those telescopes in the shadow of the
mountain. The mountain plays not only a role as a huge target, but
also a screen of cosmic rays coming into the field of view of the
detector. By mainly collecting Cherenkov photons, instead of
fluorescence photons, generated in air showers, such an array of
telescopes can work at energies as low as 1PeV, the same energy
range as the IceCube experiment. This technique is reasonably
inexpensive and the detector aperture can be easily increased in
scale to compensate for the disadvantage of this technique only
effectively detecting tau neutrinos. In the rest of this paper, we
are discussing detailed physics perspectives with one of specific
configuration of this type of detector.

Before going to more details, let us briefly review the principle
of the detection method. Incident neutrinos convert into
electrons, muons and taus in a mountain through the charged
current interaction depending upon their flavor. Electrons will
shower quickly inside the target material. Muons travel very long
distance before decaying, therefore, they are not easy to be
detected using this method. $\tau$ leptons produced inside the
mountain, that have sufficiently long life-time to escape from the
mountain, decay and induce showers in the air. Conventional air
shower detector, such as fluorescence/Cherenkov telescopes, can be
used then to detect them. The Cosmic Ray Tau Neutrino Telescope
(CRTNT) experiment is designed to detect those neutrino-induced
air showers\cite{Cao05}. To find a suitable site for the
experiment is not trivial because it must be sufficiently dry year
round and the mountain must be sufficiently steep so that there is
enough space for showers well developed to be detectable. Strongly
depending upon the specific thickness of the mountain and terrain
of the site, the CRTNT experiment needs to be reevaluated for its
feasibility with very much updated AGN neutrino models and
simulation tools. Since publication of the reference \cite{Cao05}
which addressed basic principle of the detection, there are many
progresses in AMANDAII data analysis, neutrino models and
simulation techniques. The purpose of this paper is to address
them in somewhat details.

The rest of the paper is organized as follows. The configuration
of the CRTNT detector is described in Section 2. Section 3 details
the procedure of $\tau$ neutrino converting into $\tau$ lepton in
the mountain and detecting of air shower initiated by the $\tau$
lepton. In Sections 4, 5 and 6, the event rate and sensitivity of
the CRTNT detector are estimated. We summarize all results for
this study and compare with previous works in the last section.

\section{The CRTNT Project}

The proposed CRTNT project, which currently has two telescopes as
a prototype running at Tibet site\cite{argo} and detecting cosmic
ray showers that observed by the ARGO-YBJ experiment\cite{argo},
uses fluorescence/Cherenkov light telescopes to detect air showers
induced by tau neutrinos. A candidate site is at the foothill of
Mt. Balikun, about 130 km north of Hami, Xinjiang province, China.
The contour map of the mountain is shown in figure1. The mountain
range runs in east-west direction and the northern side is quite
steep. The height of the mountain stays at about 4000 m a.s.l. for
more than 30 km in east-west direction. At the foothill, Balikun
valley stretches hundreds of kilometers at a height of
approximately 1500 m a.s.l. The total precipitation is less than
200 mm per year. The Balikun site provides an excellent convertor
for mountain-passing neutrino events and suitable weather
condition for observing the resultant air showers with
fluorescence/Cherenkov light telescopes. Preliminarily selected
sites for four CRTNT arrays, denoted as FD1, FD2, FD3 and FD4, are
located in the northern valley within 30 km from Balikun city. All
sites are convenient in terms of accessibility of power supply and
major highways. The ideal orientation of field-of-view (FOV) of
the detector enables a decent observation of the galactic center,
which is considered as the most favorable neutrino source in our
galaxy, with a considerable large exposure. An optimized
configuration of the detector array is shown in figure1.

\begin{figure}[htbp]
\centerline{\includegraphics[width=3.93in,height=2.75in]{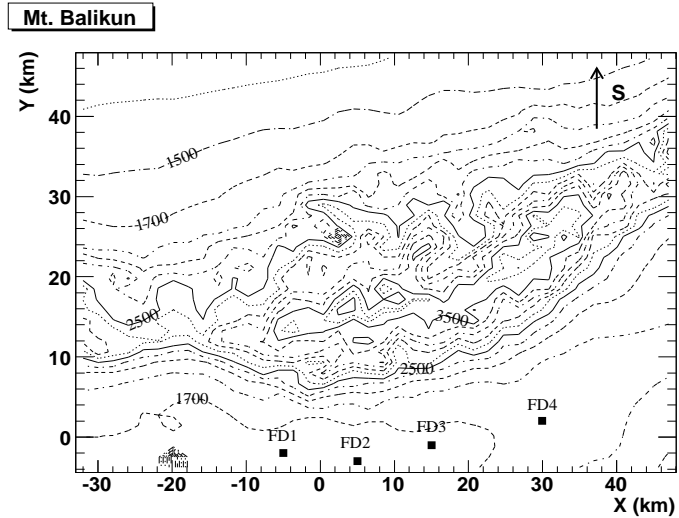}}
\label{fig1} \caption{Contour map of Mt. Balikun; the numbers
represent the altitudes in meter. Solid rectangles are the location
of four CRTNT telescope sites.}
\end{figure}

\begin{figure}[htbp]
\centerline{\includegraphics[width=3.93in,height=2.75in]{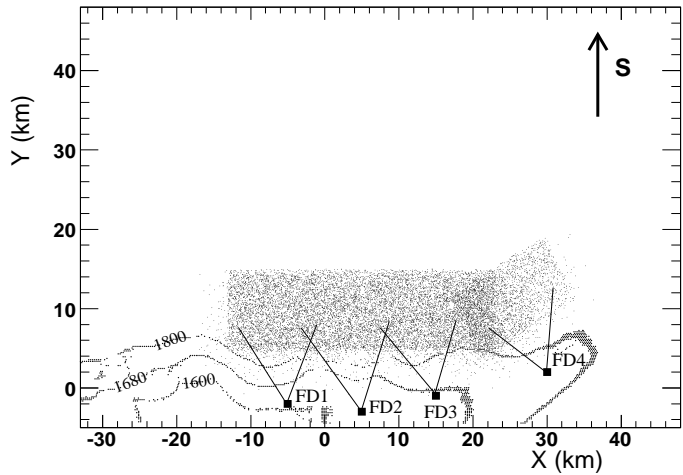}}
\label{fig2} \caption{Starting points of showers initiated by $\tau
$ leptons. Solid rectangles represent four potential CRTNT sites and
touched lines indicate the FOV of telescopes on each site. The
dotted curves show the mountain profile near the four sites.}
\end{figure}

At each site, four telescopes observe a FOV of $64\Deg$ in azimuth
angle and $14\Deg$ in elevation angle. Each telescope has a mirror
of 5 m$^{2}$ with a reflectivity of 82{\%}. A focal plane camera
consists of $16 \times 16$ photomultiplier tubes (PMTs). Each PMT
has a hexagonal photo-cathode of 44 mm from side to side with a
FOV of approximately $1\Deg \times 1\Deg$. Signals from tubes are
read out by 50MHz flash ADCs to record their complete waveforms.
An algorithm for finding pulse area is used to determine how many
photons are detected by each channel.

A successful triggered event must pass three levels of trigger
criteria. The first-level trigger is formed in every single
channel, referred to as tube trigger. It requires the
signal-to-noise ratio to be greater than $4\sigma $, where $\sigma
$ is the standard deviation of the total photon-electron noise
within a running window of 640 ns. The second-level trigger is
formed with coincidence among channels that forms certain patterns
in the camera, on the basis of a pattern recognition algorithm,
referred to as the telescope trigger. "Track-type" pattern
requires at least six triggered pixels forming a straight line
corresponding to a shower passing through the FOV. "Circular-type"
pattern requires seven triggered pixels forming a solid circle
corresponding to a Cherenkov image of a shower pointing toward the
telescope. The patterns are searched within a $6 \times 6$ box
running over the entire camera. The third-level trigger for an
event requires at least one telescope to be triggered.

\section{Monte Carlo Simulations}

Simulation of a $\tau$ neutrino event is divided into the
following three stages: 1) the $\tau$ neutrino interacts inside
the mountain; 2) the $\tau$ lepton decays and initiates an air
shower; and 3) photons are generated from the shower and detected
by the CRTNT detector. To estimate the background for the neutrino
detection, CR showers flying over the mountain are also generated.
Both the fluxes of neutrinos and CRs are assumed to be isotropic
and uniform in the FOV of the detector.

\subsection{Tau neutrino interaction}

Primary neutrinos ranging from 1 PeV to 6 EeV are sampled
according to models that predicts fluxes from active galactic
nuclei (AGN)\cite{Semikoz04,MPR2000,Stecker1992,Stecker2005}.
Incident neutrinos are uniformly distributed from $73\Deg$ to
$101\Deg$  in zenith angle and from $0\Deg$ to $180\Deg$ in
azimuth angle (the west is defined as $0\Deg$ in azimuth). All
neutrinos enter the mountain from the southern side. A
three-dimensional global coordinate system is employed to describe
incident directions, interaction positions of $\tau$ neutrinos,
and mountain profile, which are modeled by a digital topological
map. A one-dimensional coordinate system along the trajectory is
defined to describe all the three stages. If an incident
$\nu_{\tau }$ interacts inside the mountain, the energy and
momentum of the produced $\tau$ are traced until it decays. The
energy loss and the decay position of $\tau$ are simulated.
Regeneration of $\nu_{\tau}$ is taken into account, i.e. if the
$\tau $ decays inside the mountain, decay product $\nu_{\tau}$
will be traced using the same procedure again.

SHINIE (Simulation of High-energy Neutrinos Interacting with the
Earth), a Monte-Carlo simulation package, is used in description
of neutrino interaction in our simulation. In this package, the
inelasticity $y$ and the differential cross-section $d\sigma/dy$
are calculated separately, using the latest parton distribution
function CTEQ6\cite{cteq6}. The newly calculated cross section
below 1EeV is about 7\% less than the result in the package
LEPTO\cite{lepto}. Tau/muon lepton energy loss is also calculated
in details. The detailed description can be found in
\cite{Huang08} and references therein.

\subsection{Tau decay and air shower initiation}

Once a $\tau$ lepton escapes from the mountain and decays in the
air, energies of daughter particles are simulated using TAUOLA
\cite{Tauola} for all decay channels. Electrons and hadrons (the
branch ratio of 83{\%}) will initiate air showers. The air shower
carries approximately one half of the $\tau$ energy. The starting
point of shower along the particle trajectory is calculated
according to the random sampling of $\tau$ decay length, while the
shower direction is the same as that of the primary $\nu_{\tau }$.
The projection of all the starting points are shown in figure2. If
decays occur behind the detector, the shower cannot be detected
and are ignored in the simulation.

Corsika (version 6.611) \cite{Corsika} is employed to generate air
showers in the space between shower initiating points and the
CRTNT detectors. A pre-simulated shower library is established at
33 selected energies distributed between 0.1PeV and 0.6 EeV. At
each energy, 100 hadronic (pion as the primary particle) and 100
electromagnetic (electron as the primary particle) showers are
simulated. The longitudinal profiles of showers, i.e. number of
charged particles as a function of slant depth at every 5
g/cm$^{2}$ in the air, are recorded in the library. The depth of
air is calculated according to the US standard atmosphere model
(1976)\cite{atmos76}. The earth curvature is taken into account.

A shower profile is randomly selected from the library according to the
particle identification at the closest energy and the number of charged
particles in a shower is scaled up or down to represent the shower to be
simulated.

\begin{figure}[htbp]
\centerline{\includegraphics[width=3.93in,height=2.75in]{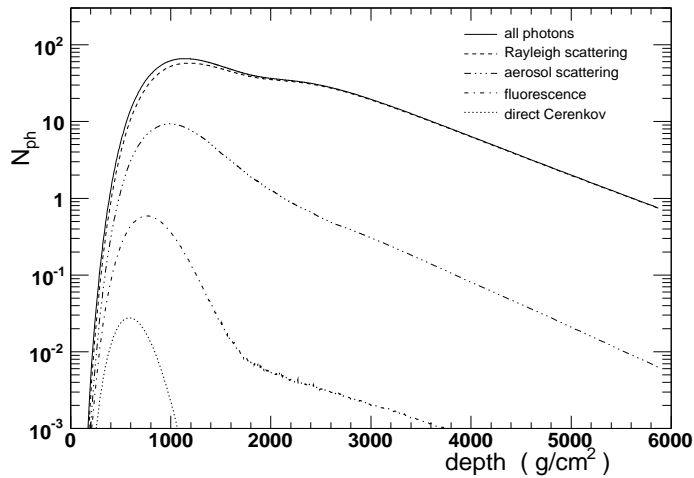}}
\label{fig3} \caption{Profiles of all photons seen by the detector
along the shower longitudinal development of a typical event. The
solid curve is the sum of all photons. The dashed curve is
Cherenkov photons scattered by atmospheric molecules (Rayleigh
scattering). The double-dot-dashed curve is Cherenkov photons
scattered by aerosols. The dot-dashed curve is fluorescence
photons. The dotted curve represents direct Cherenkov components.}
\end{figure}

\subsection{Photon production and the CRTNT detector simulation}

Ultraviolet fluorescence light is generated as charged shower
particles passing through the atmosphere. Laterally, fluorescence
photons are assumed to follow a distribution of electrons described
by Nishimura-Kamata-Greisen (NKG) formalism\cite{NKG1958}
\begin{equation}
   \rho(r) = \frac{N}{(r_{0})^2}f(s,\frac{r}{r_{0}}).
\label{eq1}
\end{equation}
where $r_{0}$ is the Moliere unit, and $s$ is the age of the
shower. The normalized function $f$ reads as
\begin{equation}
 f(s,\frac{r}{r_{0}})=(\frac{r}{r_{0}})^{s-2}(1+\frac{r}{r_{0}})^{s-4.5}\Gamma(4.5-s)/[2\pi\Gamma(s)\Gamma(4.5-2s)].
 \end{equation}

Cherenkov photons are also generated by charged particles if the
particle energy is higher than the threshold energy. Photons
scattered by the atmospheric molecules and aerosols are
distributed in all directions according to corresponding phase
functions. A standard desert aerosol model \cite{aerosol1998} with
a scale height of 1 km and a horizontal attenuation length of 25
km is assumed in the simulation. A ray-tracing procedure is
carried out to trace each photon to the photo-cathodes of PMTs
once photons are generated. All detector responses are considered
in the ray-tracing procedure. A detailed description of the
ray-tracing procedure can be found elsewhere \cite{Hires01a} and
references therein. In figure3, profiles of all kinds of photons
that collected by the CRTNT detector are plotted as functions of
slant atmospheric depth along the shower longitudinal development
for a typical simulated shower.

All photons collected by one PMT form a complete waveform
according to their arrival time. Night sky background (NSB)
photons with a flux of 40 photons per microsecond per square meter
of the light collector are randomly added to the waveform. The
electronic noise with a mean of 1.2 FADC counts per 20 ns is also
added to every channel. Triggering algorithm at three levels as
described in Section 2 is implemented in the simulation. In
figure4, an example of a neutrino-induced shower event detected by
CRTNT is plotted.

\begin{figure}[htbp]
\centerline{\includegraphics[bb=9mm 39mm 205mm
125mm,width=4.13in,height=1.50in]{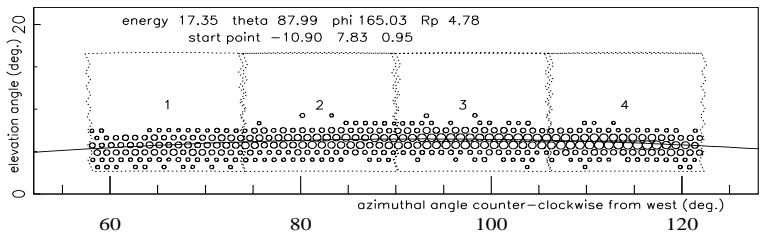}} \label{fig4}
\caption{ A typical horizontal air shower seen by the CRTNT
detector. Each square marked with a number shows the FOV of each
telescope. Circles represent triggered tubes and the size of each
circle is proportional to logarithm of the number of photons. The
solid line represents a plane containing shower axis and the
detector. Energy (in PeV), zenith and azimuth angle (in degree),
impact parameter and coordinates of shower starting point (in km)
are displayed.}
\end{figure}

\subsection{Cosmic ray background simulation}

As a background, CR showers are simulated with core location
uniformly distributed in an area of 32 km $\times $ 10 km between
the mountain and detectors. CR shower energies are randomly
selected from a spectrum of $E^{-3}$ between 3 PeV to 1 EeV. All
showers are between $70\Deg$ and $75\Deg$ in zenith angle and
between $0\Deg$ and $180\Deg$ in azimuth angle. Cosmic rays from
smaller zenith angles ($\theta < 70\Deg$) can be rejected
according to their reconstructed geometry without any ambiguity.
CRs from a larger zenith angles ($\theta > 75\Deg$) are screened
by the mountain.

Using Corsika (version 6.611), a shower library similar as that in
neutrino simulation at 26 selected energies distributed from 3PeV
to 1EeV and zenith angle at 75 degrees is established. At each
energy, 100 hadronic showers with proton as primary particles are
produced.

In the simulation, cosmic ray showers are randomly selected from
the shower library. The slant atmospheric depth with an earth
curvature correction, photon production, light propagation and
trigger algorithm are the same as those used in the neutrino
simulation. A uniform mountain profile with a height of 2.5 km is
assumed to be 8 km away from the detectors serving as a screen.

\section{Estimation of event rate}

Using the algorithm described in Section 3, a
$\nu_{\tau}$-to-shower conversion efficiency of $1.92 \times
10^{-4}$ is yielded. An average trigger efficiency of showers
induced by the products of $\tau$ decay is found to be
approximately 21.8{\%}. The input and observed event distribution
is shown in figure5. By using the AGN neutrino flux proposed by
\cite{Semikoz04}, the event rate is 35.7 per year assuming a duty
cycle of 15{\%} for the CRTNT detector. Although this AGN flux
model was ruled out by the AMANDA experiment \cite{AMANDA07} we
still use it for comparison with our previous study \cite{Cao05},
where the same model was used.

\begin{figure}[htbp]
\centerline{\includegraphics[width=3.93in,height=2.75in]{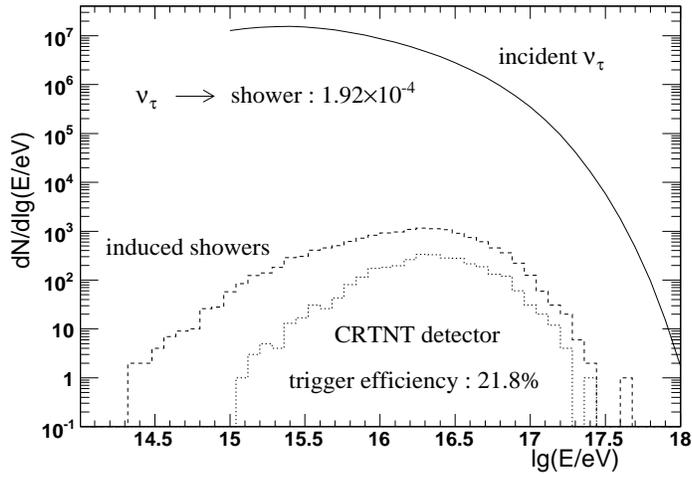}}
\label{fig5} \caption{Input and observed event distribution using
CRTNT array in Mt. Balikun. The solid curve represents the
incident neutrino spectrum, the dashed curve is the induced shower
energy distribution, and the dotted curve shows the triggered
events energy distribution.}
\end{figure}

We also estimate the event rates of other two AGN models. The
$\nu_{\tau}$ to $\tau$ conversion efficiency of MPR AGN jet
model\cite{MPR2000} is $2.14\times10^{-4}$. The triggering
efficiency is $23.5\%$. The expected event rate is 27.3 per year.
For SDSS AGN core model\cite{Stecker1992, Stecker2005}, the
$\nu_{\tau}$ to $\tau$ conversion and triggering efficiency are
$7.25\times10^{-5}$ and $12.5\%$, respectively. The expected event
rate is 5.9 per year for this model.

 Event rate of CR showers is estimated using average flux
$J(E)= 2 \times 10^{24}\times E^{3}(eV^{2}\cdot m^{-2}\cdot
s^{-1}\cdot sr^{-1}) $ \cite{Hires01b}. The trigger efficiency is
found to be about 0.29\% for CR showers. The input and observed
spectra are shown in figure6. In the simulation corresponding to
an exposure of more than two years, 93407 CR events are collected,
i.e. 38300.7 events will be detected per year.

\begin{figure}[htbp]
\centerline{\includegraphics[width=3.93in,height=2.75in]{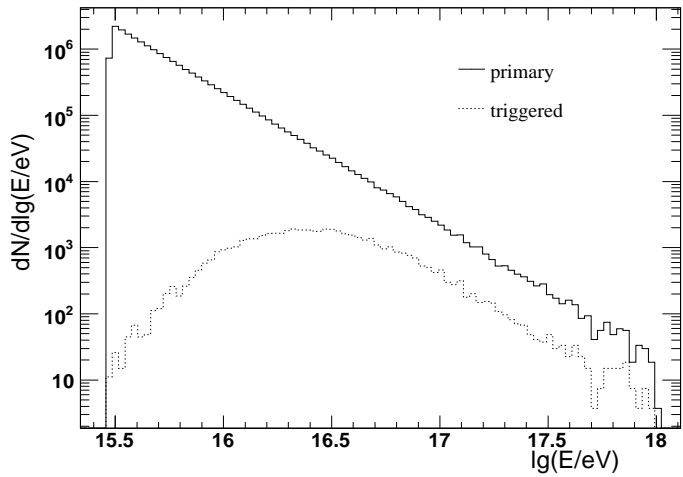}}
\label{fig6} \caption{Energy distribution of primary CRs (solid
curve) and triggered events (dashed curve).}
\end{figure}

\section{Event selection}

Neutrinos and CRs produce showers with distinct characteristics
because they come from different directions and develop in
different depths of the atmosphere. Without detailed shower
reconstruction, a simple algorithm can identify neutrino events
from CR background events by sorting potential neutrino events
into the following five types.

\begin{enumerate}
\item {\textbf{Up-going event} \\
A shower detector plane (\textbf{SDP}) is defined as a plane that
contains the shower axis and the detector. An apparent up-going
event, caused by neutrinos, can be identified by finding the
elevation angles of triggered tubes along the SDP increasing with
time. Only by checking on the shower flying direction, a cut
criterion in time difference between the highest and the lowest
tubes, denoted as $dt$, will pick out the majority of neutrino
events (41.5{\%}). The $dt$ distributions for apparent up-going
neutrino ($dt>0$) and all CR showers are shown in figure7(a). In
order to avoid cosmic ray showers that apparently go upwards, a
cut at 200ns is chosen.}

\item {\textbf{Horizontal event} \\
An event whose SDP normal vector lies within a cone of 5 degrees
from zenith is called a horizontal event. Such events might not be
picked up as neutrino events by timing (i). The space angle
between SDP normal vector and the zenith, denoted as $\psi$,
distributes as in figure7(b). After the above step/s (the same in
the following), approximately 7.9{\%} of neutrino events are of
this type. No cosmic rays belong to this category.}

\item {\textbf{Neutrino induced Cherenkov event} \\
Head-on showers generate Cherenkov images concentrating in an
ellipse-like pattern. Because all photons arrive to the detector
almost at the same time, timing criterion does not work for this
type of events. However, a neutrino induced Cherenkov event must
have the elevation angle of the center of image \textbf{(COM)},
defined as a mean position weighted by triggered PMT signals,
lower than $11\Deg$, because they come out from the mountain. CR
events from such low directions start at very far away and are
blocked by mountains. Only small portion of them which just fly
over the tip of the mountain may generate similar patterns, but
the COMs must be at high positions. Figure7(c) shows distributions
of elevations of the COMs for different showers. A cut at $11\Deg$
makes a clear separation between them. Approximately 17.7{\%} of
neutrino events belongs to this category. }

\item {\textbf{Very long track event} \\
Once an energetic shower starts at outside the FOV of the
detectors, their shower tracks could be observed by multiple
mirrors at one site as a characteristic. If it is induced by a
mountain-passing neutrino, it is once again characterized by its
low elevation of the COM. Therefore, cutting on both COM elevation
and number of rows of triggered tubes in a shower as shown in
figure7(d), i.e. the COM elevation must be lower than 10$\deg$ and
number of rows must be less than 13, neutrino events will be
picked out. Only 7.2{\%} events among all neutrino events belong
to this category.}

\item {\textbf{Back-to-mountain event} \\
There must be some events with very clear characteristics that the
first triggered tube located in the central area of the FOV of a
telescope, thus strongly indicate that $\tau$ leptons come out
from the mountain and initiate a shower start from where the tube
points at. Without many efforts, those neutrino events should be
picked out if shower images start from points that are certainly
not associated with any edges of the FOV. CR shower images must
engaged with the edges because the FOV is screened by the
mountain. As a quantitative criterion, 4 $\deg$ along SDP between
the start point of an image to one of the edges is applied to
select this type of events. This takes the reconstruction error of
SDP into account. However, only 5.8{\%} of neutrino events are
falling in this category.}

\begin{figure}[htbp]
\begin{center}
\scalebox{0.3}{\subfigure[]{\includegraphics{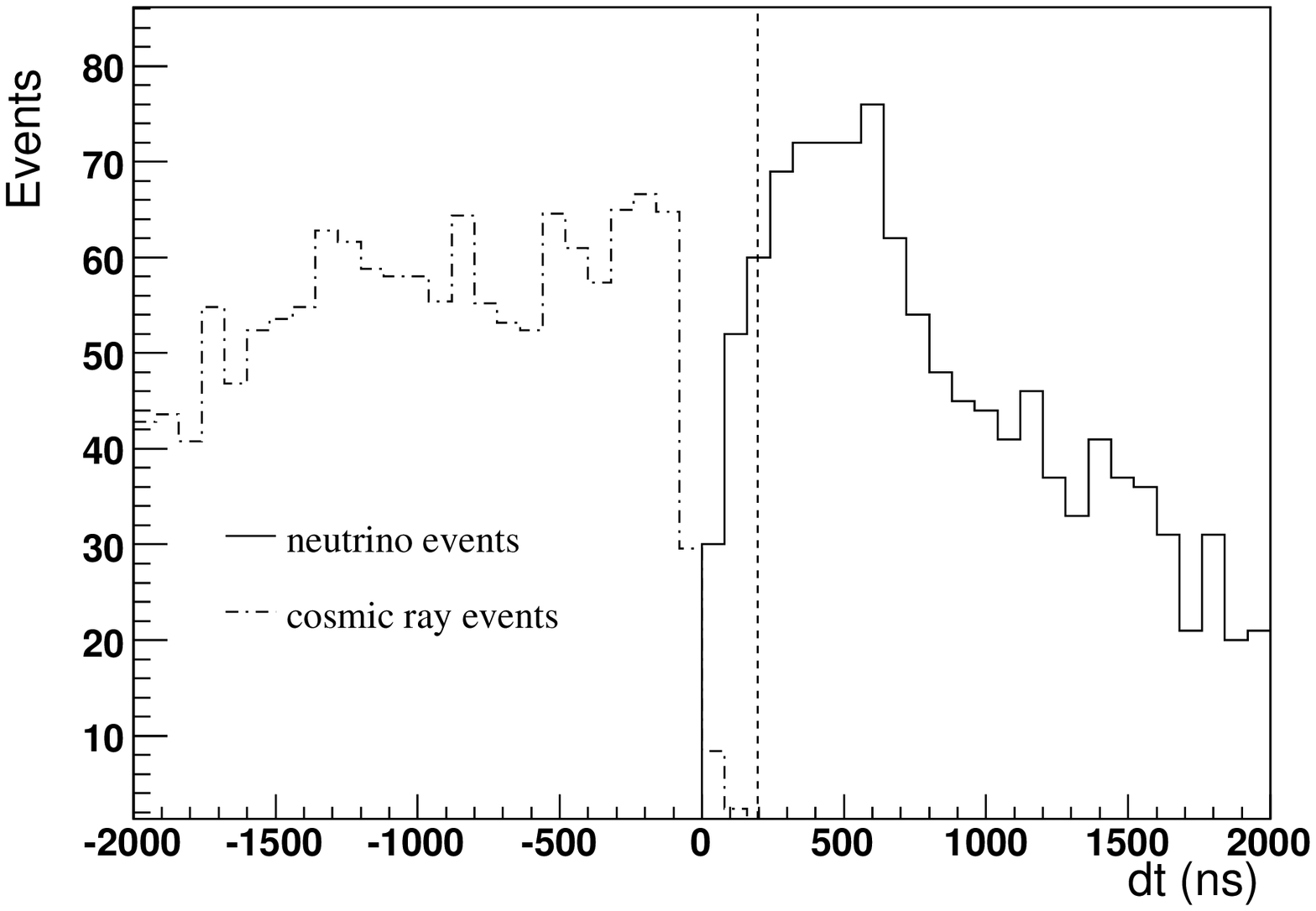}}}
\scalebox{0.3}{\subfigure[]{\includegraphics{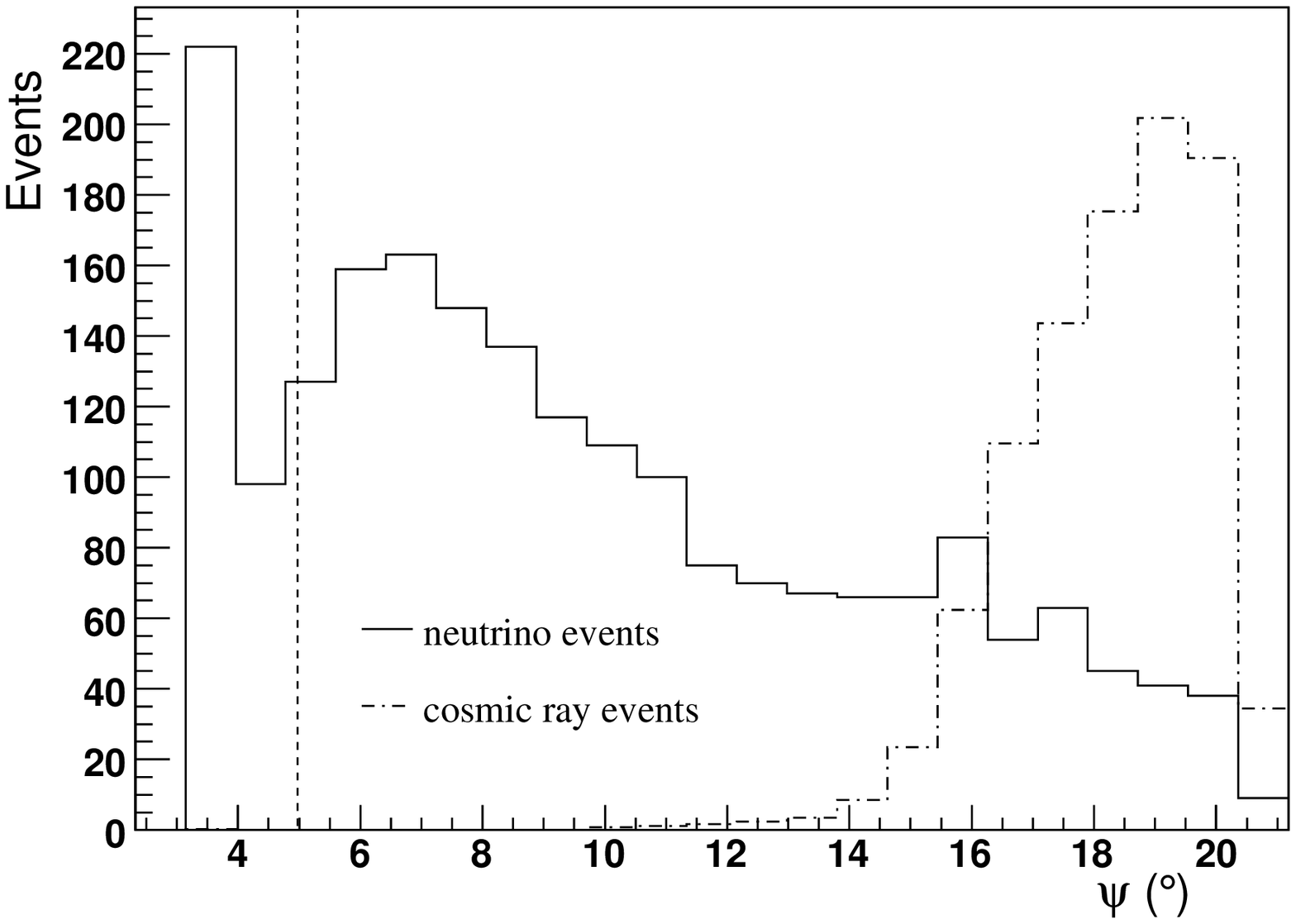}}}
\scalebox{0.3}{\subfigure[]{\includegraphics{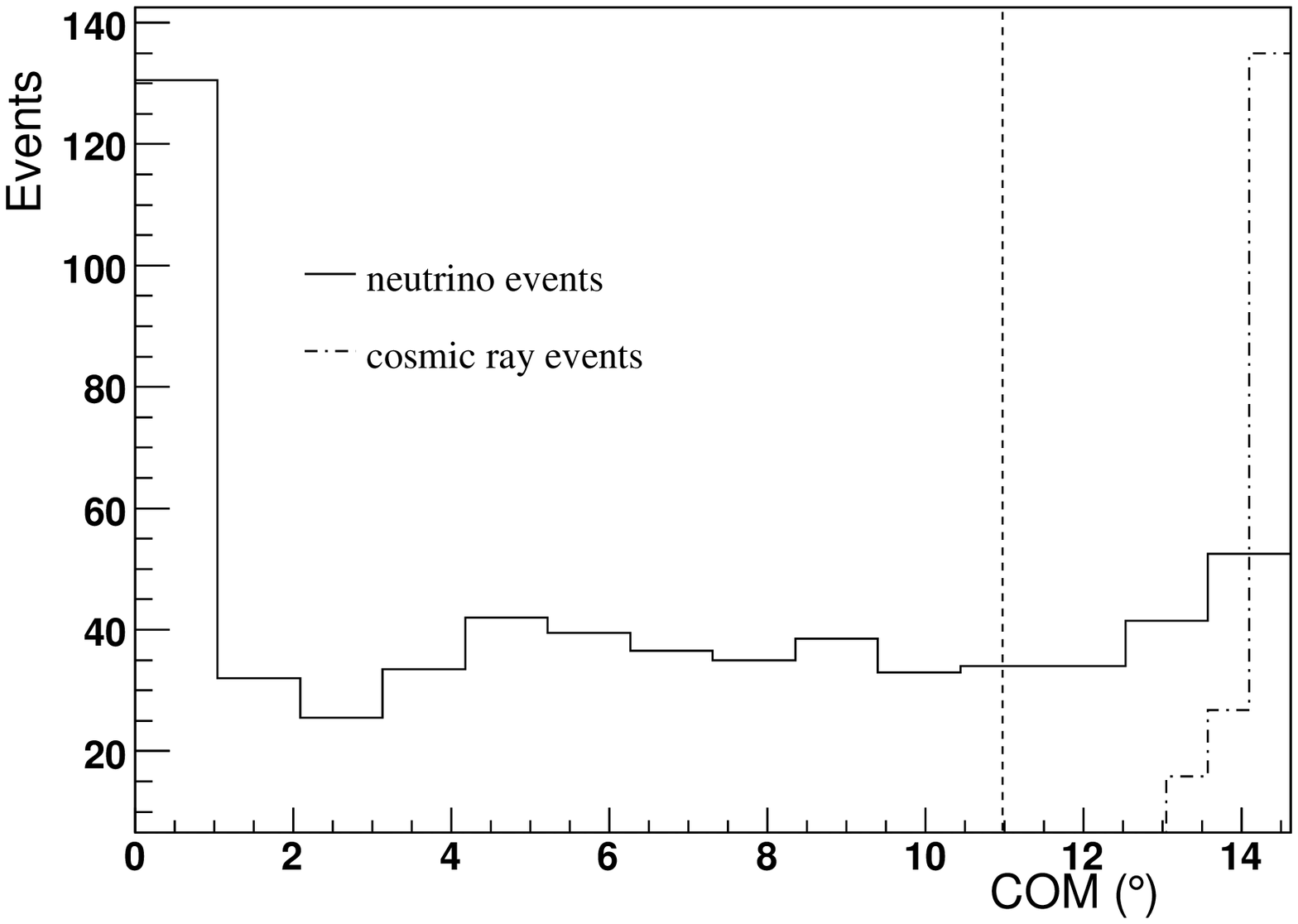}}}
\scalebox{0.3}{\subfigure[]{\includegraphics{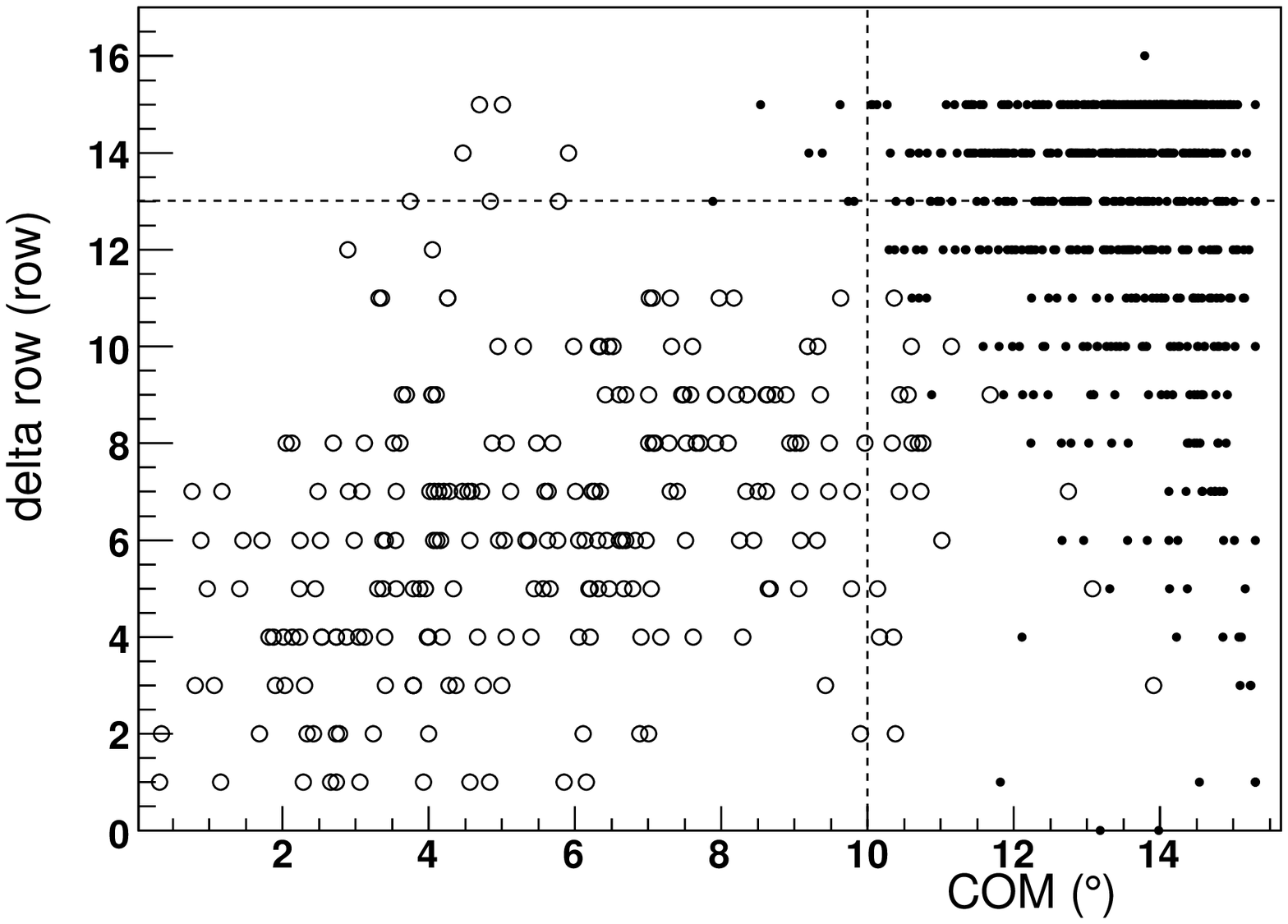}}}
\label{fig7} \caption{Parameters distribution for neutrino induced
events (solid curves in figure (a), (b) and (c), open circle in
figure (d)) and cosmic ray induced events (dash-dotted curves in
figure (a), (b) and (c), solid circle in figure (d)). The dashed
lines represent the cut criteria. (a) Average flying time for a
shower over the FOV of CRTNT detector. (b) The distribution of the
space angle between SDP normal vector and zenith. (c) COM
elevation distribution for the cherenkov events. (d) Number of
rows of shower images versus elevation of the center of image for
very long track events. (see text for more details.)}
\end{center}
\end{figure}
\end{enumerate}

Summing them up, 80.1{\%} of neutrino showers can be identified.
The event rate for AGN(Semikoz et al., 2004) reduces to 28.6 per
year. The event rates for MPR AGN jet model and SDSS AGN core
model are 21.9 and 4.7 per year, respectively.

To estimate how many CR showers are mis-identified as neutrino
events, we apply the proposed event selection algorithm to 93407
CR events, and only one of them is selected as a neutrino event.
This yields an average CR background of 0.4 event per year for the
neutrino detection.

\section{Sensitivity}
We estimate the sensitivity of CRTNT project using modified
Feldman-Cousins method\cite{Hill2003}. The survival probability of
triggered neutrino events, defined as the number of identified
neutrino events out of the total, is used as a reference parameter
in the optimization procedure. By tuning the criteria, the
survival probability changes from 67\% to 92\% as listed in Table
1. The minimum upper limit goes to 19.9 $(eV\cdot cm^{-2}\cdot
s^{-1}\cdot sr^{-1})$ with a hypothetical neutrino flux
$\Phi_{s}=E^{2}\cdot J(E)=10^{2} (eV\cdot cm^{-2}\cdot s^{-1}\cdot
sr^{-1})$ with energies from $10^{15} eV$ to $10^{18} eV$.

\begin{table}
\caption{Optimization procedure of neutrino event selection.
}\label{table:1}
\begin{center}
\begin{tabular}{ccccc}
\hline
 $R(\%)$ & $n_{s}$ & $\langle n_{b}\rangle$ & $\overline{\mu_{90}}$ & upper limit\\\hline
 67.4     &11.73      &0.00     &2.44     &20.8 \\
 80.1     &13.94      &0.41     &2.78     &19.9 \\
 85.3     &14.84      &0.80     &3.03     &20.4 \\
 92.0     &16.08      &1.23     &3.46     &21.5 \\\hline
\end{tabular}
\end{center}
{\it R is the neutrino event survival probability. $n_{s}$
represents the expected number of events per year from the model
of $\Phi_{s}$ after the event selection. $\langle n_{b}\rangle$ is
the corresponding number of background cosmic ray events per year.
$\overline{\mu_{90}}$ shows the average upper limit from
Feldman-Cousins method\cite {Feldman98}. The upper limit is in
unit of $eV\cdot cm^{-2}\cdot s^{-1}\cdot sr^{-1}$}.
\end{table}

As the result, if no excess is observed in one year, an upper
limit of $19.9(eV\cdot cm^{-2}\cdot s^{-1}\cdot sr^{-1})$ can be
set with 90\% C.L. by the CRTNT experiment. It falls to $6.7
(eV\cdot cm^{-2}\cdot s^{-1}\cdot sr^{-1})$ for three years
observation. Figure8 shows the corresponding results comparing
with other experiments. As references, three neutrino source
models together with atmospheric and cosmogenic neutrino models
are plotted in the same figure as well.

\begin{figure}[htbp]
\centerline{\includegraphics[width=3.93in,height=2.75in]{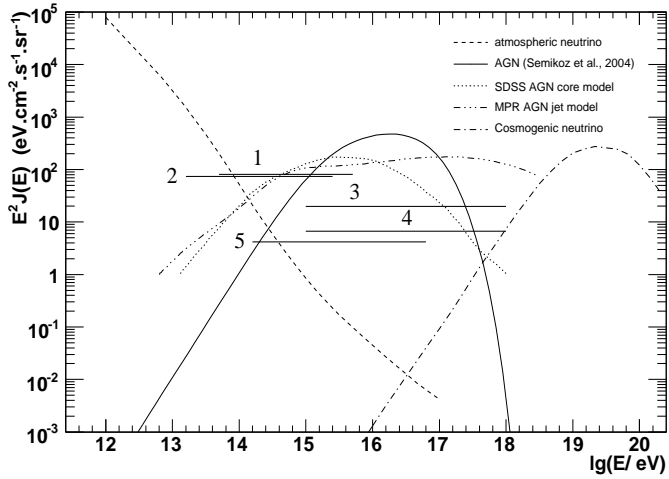}}
\label{fig8} \caption{Predicted diffuse neutrino fluxes and
sensitivities of CRTNT project (horizontal solid lines) for tau
neutrinos and other three experiments for muon neutrinos. The
lines marked with numbers represent the upper limit of different
experiments. Line 3 and Line 4 are for CRTNT with one year and
three year of observation, respectively. Line 1 is for Antares in
one year of observation\cite{Antares05-1}. Line 2 is for Amanda II
within 807 days observation\cite{AMANDA07}. Line 5 is for IceCube
with three year of observation\cite{Icecube04}.}
\end{figure}

\section{Conclusions}

A complete simulation chain is developed including neutrino
interaction, $\tau$ lepton decay, air shower and light production,
detector responses, triggering, and neutrino event selection
algorithm. The event rate for a proposed AGN neutrino flux
(Semikoz et. al, 2004) is found to be 28.6 per year. The event
rates for MPR AGN jet model and SDSS AGN core model are 21.9 and
4.7, respectively.

Comparing with our previous estimate for the Mt. Wheeler site
\cite{Cao05}, the expected annual event rate increases from 8 for
three sites at Mt. Wheeler Peak to 28.6 for four sites at Mt.
Balikun site according to the same AGN model. The significant
improvement comes from 1) a much larger FOV of the CRTNT detector
array at the Mt. Balikun site because the mountain stretches
longer with a height about 4 km a.s.l.; 2) an improvement of the
detector design by widening the bandwidth of light signals which
are composed mainly of scattered Cherenkov light distributing over
a range up to 600 nm; and 3) a correction of an error in the
previous simulation code about Rayleigh scattering of Cherenkov
photons which caused an underestimation of shower trigger
efficiency by a factor of 2.5. The overall effect is about an
enhancement of the event rate by a factor of 4.5. Other
improvements have been made in the simulation by using more
detailed description about neutrino interaction, $\tau$ lepton
propagation in the mountain and its decay. Air shower generation
is also improved as well.

According to a parametrization of CR event distribution, isotropic
cosmic rays yield a background of 38300.7 events per year in a
zenith angle ranging from 70\Deg to 75\Deg. By using the neutrino
event selection algorithm, 80.1{\%} of 35.7 neutrino events with
0.4 CR background event per year are picked out. If CRTNT does not
see any signal excess with one year of observation, an upper limit
of $19.9 (eV\cdot cm^{-2}\cdot s^{-1}\cdot sr^{-1})$ with 90{\%}
C.L. can be concluded.

\ack{This work and CRTNT project is supported by the Innovation
fund (U-526)of the Institute of High Energy Physics (IHEP) and
Hundred Talents \&\ Outstanding Young Scientists Abroad Program
(U-610/112901560333) from IHEP and Chinese Academy of Sciences.
Authors MAH and TCL are supported by National Science Council of
Taiwan under project number NSC-95-2112-M239-003 and
NSC-96-2112-M239-001.}

\end{document}